\documentclass[%
 aip,
 jmp,%
amsmath,amssymb,
 reprint,%
]{revtex4-1}
\usepackage{amsmath}

\usepackage{amssymb}
\usepackage[permil]{overpic}
\usepackage{graphicx}
\usepackage{dcolumn}

\begin{document}

 \preprint{AIP/123-QED}

\title{Optical responses of the switching currents in Al and Nb Josephson junctions}

\author{Yiwen Wang}
 \affiliation{
Quantum Optoelectronics Laboratory, School of Physics, Southwest
Jiaotong University, Chengdu 610031, China}

\author{Pinjia Zhou}
 \affiliation{
Quantum Optoelectronics Laboratory, School of Physics, Southwest
Jiaotong University, Chengdu 610031, China}

\author{Lianfu Wei}
 \email{weilianfu@gmail.com}
 \affiliation{
Quantum Optoelectronics Laboratory, School of Physics, Southwest
Jiaotong University, Chengdu 610031, China}
 \affiliation{
State Key Laboratory of Optoelectronic Materials and Technologies,
School of Physics and Engineering, Sun Yat-Sen University, Guangzhou
510275, China}

\author{Beihong Zhang}
 \affiliation{
Quantum Optoelectronics Laboratory, School of Physics, Southwest
Jiaotong University, Chengdu 610031, China}

\author{Qiang Wei}
 \affiliation{
Quantum Optoelectronics Laboratory, School of Physics, Southwest
Jiaotong University, Chengdu 610031, China}

\author{Jiquan Zhai}
 \affiliation{
Research Institute of Superconductor Electronics, Department of
Electronic Science and Engineering, Nanjing University, Nanjing
210093, China}

\author{Weiwei Xu}
 \affiliation{
Research Institute of Superconductor Electronics, Department of
Electronic Science and Engineering, Nanjing University, Nanjing
210093, China}

\author{Chunhai Cao}
 \affiliation{
Research Institute of Superconductor Electronics, Department of
Electronic Science and Engineering, Nanjing University, Nanjing
210093, China}

\begin{abstract}
We experimentally demonstrated the optical responses of the
switching currents in two types of Josephson tunnel junctions:
Al/AlOx/Al and Nb/AlOx/Nb. The radiation-induced switching current
shifts were measured at ultra-low bath temperature ($T$ $\approx$
$16$ mK). It is observed that the Al-junction has a more sensitive
optical response than the Nb-junction, which is as expected since Al
electrode has a smaller superconducting gap energy. The minimum
detectable radiation powers with the present Al-junction and
Nb-junction are $8$ pW (corresponding to $8$ $\times 10^{5}$
incoming photons in one measurement cycle) and $2$ nW respectively.
In addition, we found that the radiation-induced thermal effects are
dominant in the observed optical responses. Several methods are
proposed to further improve the optical responsivity, so that the
josephson junction based devices could be applicable in photon
detections.
\end{abstract}

\maketitle

Superconducting photon detectors at near-infrared wavelengths, with
photon-number resolving power, have shown great promises in quantum
optics and quantum information applications. The superconducting
detectors mainly include superconducting nanowire detectors
(SNSPDs)~\cite{G,H,A,AA}, transition-edge sensors
(TESs)~\cite{Miller,Miller2}, superconducting tunnel junctions
(STJs)~\cite{SF,JD} and microwave kinetic inductance detectors
(MKIDs)~\cite{pkday1,gao1,gao2}, etc.. Actually, photon detections
can also be achieved through other ways, such as by measuring the
changes in the critical current of a Josephson tunnel junction due
to radiation. Physically, a photon with sufficient energy $h\nu$
($>$ $2\Delta$) can directly break $\eta h\nu/2\Delta$ Cooper pairs,
where $\Delta$ is the superconducting gap energy and $\eta$ the
convert efficiency. Therefore, when a photon is incident on one
superconducting electrode of the Josephson junction, excess
quasiparticles will be excited and the Cooper pair density on the
irradiated electrode will decrease. This will lead to an abrupt
reduction in the critical current $I_{c}$ since~\cite{feynman}
$I_{c}\propto \sqrt{\rho_{1}\rho_{2}}$, where $\rho_{i}$ is the
Cooper pair density on the $i$-th electrode. On the other hand,
phonons in the substrate around the radiation center may be excited
and thus cause a temperature increase nearby the junction area. This
thermal effect can also reduce the critical current $I_{c}$ based on
the relation~\cite{AB}
\begin{equation}
I_{c}R_{n}=[\pi\Delta(T)/2e]\tanh[\Delta(T)/2k_{B}T],
\end{equation}
with $R_{n}$ being the normal state resistance and $T$ the bath
temperature. Since the gap energy $\Delta(T)$ decreases with $T$,
the critical current $I_{c}$ also decreases with $T$. Therefore,
both pure pair-breaking effects and thermal effects can lead to a
reduction in the critical current. This provides a feasible way to
detect the incident photons via measuring the radiation-induced
changes in the critical current of a Josephon junction.

Note that the well-known $ac$ Josephson effect was utilized to
detect the microwave and far-infrared radiation several years
ago~\cite{shapiro}. Later, the superconducting gap voltage shifts
due to visible and infrared radiation were measured in Nb/AlOx/Nb
junctions~\cite{Osterman1} and junction arrays~\cite{Osterman2} at
temperatures around $4.2$ K. Specifically, the junctions immersed in
superfluid helium were observed to have lower optical responsivity
compared to those in vacuum. This is because the heating effect is
suppressed in liquid helium and thus the optical responses of the
devices are entirely due to the pair-breaking mechanism. Other
experiments~\cite{wire,Mont1,Mont2,Mont3} with Nb junctions had also
verified these responses due to pure pair-breaking and thermal
effects.

However, the previous experiments were all done with Nb junctions.
In our experiments, we studied the optical responses of both Al- and
Nb junctions. We found that the Al-junction has a more sensitive
optical response than Nb-junction. This is a reasonable observation
since aluminum has a smaller gap energy. Thus, a certain radiation
energy can break more Cooper pairs on Al electrode. Besides, in all
of the previous experiments the Josephson junctions were biased at
constant currents and the gap voltage shifts were measured as the
optical responses. Alternatively, we swept the bias current through
the junction and measured the switching current responses to a
continuous radiation at $1550$ nm. This detection approach is
relatively simple and has not been reported before, as far as we
know. Moreover, the previous experiments were all done at
temperatures around $1$ K $\sim$ $4.2$ K while our system works in
an ultra-low temperature regime, i.e., the bath temperature $T$
$\approx$ $16$ mK. Thermal noise in the circuit is minimized at such
low temperatures and thus devices are expected to have more
sensitive optical responses.

For our measurements, the Al/AlOx/Al junction was fabricated by
electron beam double-angle evaporation, and the Nb/AlOx/Nb junction
was fabricated by magnetron sputtering and ion etching. For both Al
and Nb devices, the junction areas are about $6$ $\mu m^{2}$ and the
top electrodes exposed for illumination are about $100$ nm thick.
The chips are cut to approximately $2$ mm $\times$ $2$ mm, with
Si-substrates of $0.5$ mm thick. Both junctions are slightly
damped~\cite{Stewart} and show hysteretic IV curves with small
retrapping currents and sharp onsets of voltage at the maximum bias
currents (i.e., the switching currents).
%
%
%
\begin{figure}[tbp]
\begin{center}
\begin{overpic}[width=7.6cm]{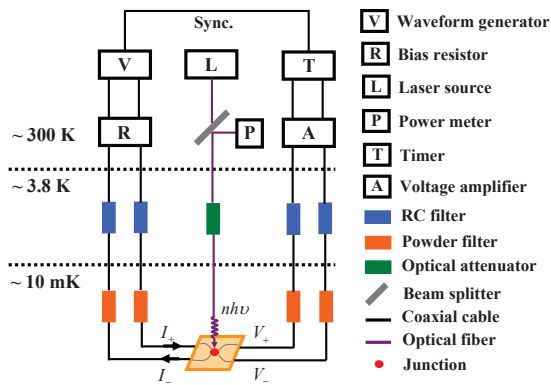}
\end{overpic}
\end{center}
\small {\caption{(Color online)  Schematics of the measurement
setup. Four-probe method is used to measure the junction
current-voltage characteristics. The tested junction is placed in
the sample cell at $\sim$ $16$ mK and irradiated by $1550$ nm laser
beam.}}
\end{figure}
%

%

The schematics of our measurement setup are shown in Fig.~1. The
measured junction is placed in a superconducting aluminum sample
cell, mounted at the mixing chamber in a dilution refrigerator with
base temperature around $10$ mK. Four-probe technique is used to
measure the current-voltage characteristics of the devices. The
waveform generator can output a voltage signal, which is applied to
a resistor to generate a bias current through the junction. The
voltage response is amplified by a battery-powered pre-amplifier and
then fed into a timer. All electrical leads, connecting the sample
cell to room temperature electronics, are filtered by low-pass RC
filters and copper powder microwave filters. To radiate the
junction, a single-mode optical fiber is set up from the room
temperature environment down to the sample cell. A laser source is
connected to the top end of the fiber and generates a steady
radiation of wavelength $1550$ nm. The bottom end of the fiber is
carefully aligned and fixed, so that the laser beam can focus on the
top electrode of the junction. The fiber end is estimated to be
about $200$ $\mu$m vertically away from the chip surface and the
irradiated area is about $80$ $\mu$m in diameter. Therefore, the
junction area is completely covered by light.

Due to the presence of thermal fluctuations and quantum tunneling,
the junction switches from the zero-voltage state to the finite
voltage state at a bias current $I_{s}$ smaller than its critical
current $I_{c}$. Since this switching is a deterministic random
process, the switching current $I_{s}$ shows a Lorentzian
distribution~\cite{Fulton,Barone}, which can be mainly characterized
by the width $\sigma_{s}$ and mean value $\langle I_{s}\rangle$. In
our experiment the switching current distribution $P(I_{s})$ is
measured by using the time-of-flight method~\cite{yiwen}. For each
switching event, the bias current is ramped linearly from a value
below zero up to a value higher than the critical current $I_{c}$.
When the junction switches from the zero-voltage state to the
finite-voltage state, the timer will be triggered to record the
switching time and the corresponding switching current $I_{s}$ can
be calculated from the ramping rate. The bias current is then
reduced to below zero, resetting the junction to the zero-voltage
state. The repetition frequency is $71.3$ Hz and the measurement
cycle is repeated $2\times10^{3}$ times to obtain an ensemble of
$I_{s}$, from which the distribution of switching current $P(I_{s})$
can be obtained.
\begin{figure}[tbp]
\begin{center}
\begin{overpic}[width=7.3cm]{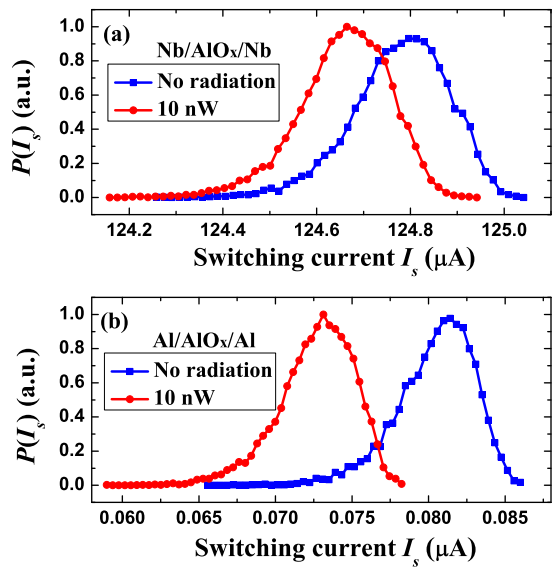}
\end{overpic}
\end{center}
\small {\caption{(Color online) (a) The measured distributions of
the switching currents of Nb/AlOx/Nb. Blue squares correspond to the
data in the absence of radiation and red dots correspond to a $10$
nw radiation on the junction. (b) The switching current
distributions of Al/AlOx/Al without radiation and with a $10$ nw
radiation respectively.}}
\end{figure}

Fig.~2 plots the measured switching current distribution, i.e., the
switching probability $P(I_{s})$ as a function of the switching
current $I_{s}$. Fig.~2(a) shows the switching current measurements
on Nb/AlOx/Nb at temperature $16$ mK. The blue curve is the
distribution in the absence of radiation, from which one can
calculate the mean switching current $\langle I_{s0}(Nb)\rangle $ =
$124.78$ $\mu$A and the distribution width (standard deviation)
$\sigma_{s0}(Nb)$ = $102.73$ nA. The red curve corresponds to a $10$
nW radiation on the junction electrode. In this case the mean
switching current shifts down to $\langle I_{s}(Nb)\rangle $ =
$124.66$ $\mu$A and the distribution width $\sigma_{s}(Nb)$ =
$98.00$ nA. Fig.~2(b) shows the same measurements on Al/AlOx/Al
under the same experimental conditions. Without radiation, the mean
switching current is $\langle I_{s0}(Al)\rangle $ = $80.44$ nA and
the distribution width $\sigma_{s0}(Al)$ = $2.53$ nA. In the
presence of $10$ nW radiation, the mean switching current is
$\langle I_{s1}(Al)\rangle $ = $72.65$ nA and the distribution width
$\sigma_{s1}(Al)$ = $2.47$ nA.

There are two ways to define the photon responsivity of the device.
One is the ratio of the response to noise, i.e., $R_{a}$ $=$ $\Delta
I_{s}/\sigma_{s0}$ $=$ $(\langle I_{s0} \rangle -\langle I_{s1}
\rangle)/\sigma_{s0}$. By this way, we have $R_{a}$ $=$ $1.17$ for
the Nb-junction and $R_{a}$ $=$ $3.08$ for the Al-junction, showing
that the Al device has a higher photon responsivity. The second way
to define photon responsivity is the relative shift of switching
current, i.e., $R_{b}$ $=$ $\Delta I_{s}/I_{s0}$ = $(\langle
I_{s0}\rangle - \langle I_{s1}\rangle)/\langle I_{s0}\rangle$. In
this way, we obtain $R_{b}$ = $9.6$ $\times 10^{-4}$ for the
Nb-junction and $R_{b}$ $=$ $9.7$ $\times 10^{-2}$ for the
Al-junction, showing again that the Al device has a more sensitive
response. We take the second definition of responsivity in the
following discussions.

\begin{figure}[tbp]
\begin{center}
\begin{overpic}[width=7.5cm]{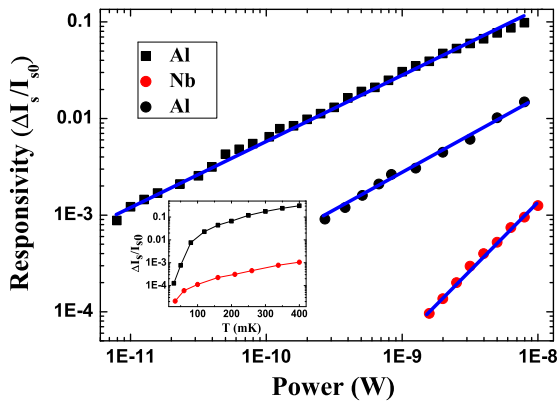}
\end{overpic}
\end{center}
\small {\caption{(Color online) The logarithmic $\Delta
I_{s}/I_{s0}$ as a function of logarithmic radiation power. Black
squares and black circles are the experimental data for focusing the
light on junction and nearby substrate of Al device respectively.
Red circles correspond to radiation on the junction of Nb device.
The blue lines are the linear fitting, from which one can obtain the
power law dependence of the responsivity. The inset shows $\Delta
I_{s}/I_{s0}$ as a function of the bath temperature.}}
\end{figure}

We now investigate the optical responses of the switching currents
under different radiation powers. To this aim we varied the light
intensity and measured the corresponding average switching current
at the base temperature $T$ = $16$ mK. Fig.~3 shows the relative
switching current shift $\Delta I_{s}/I_{s0}$ (i.e., the
responsivity $R_{b}$) as a function of the radiation power. The
black squares and red circles correspond to irradiations on Al- and
Nb-junction respectively. It is shown that the logarithmic switching
current shift increases linearly with the logarithmic radiation
power in the applied power range. By fitting the line slope, one can
find that $R_{b}$ is approximately proportional to $P^{0.6}$ for the
Al-junction while proportional to $P^{1.2}$ for the Nb-junction. The
minimum radiation power that the Al device can detect is about $8$
pW (corresponding to $8$ $\times 10^{5}$ incoming photons per
measurement cycle), which is much smaller than the minimum power of
$2$ nW that Nb device can detect. The device response to low
radiation power is limited by the average switching current
fluctuations, which are mainly due to the inevitable low-frequency
noises in the electronics.

The switching current shift increases with the radiation power,
which is qualitatively similar to its bath temperature dependence.
The inset of Fig.~3 shows $\Delta I_{s}/I_{s0}$ as a function of the
bath temperature in the range of $30$ mK to $400$ mK, where the
measured switching current shift increases with temperature. This
suggests that the thermal effects are dominant in the observed
radiation power dependence of $R_{b}$ at $T =$ $16$ mK.
Experimentally, most of the photons are incident on the substrate
rather than the superconducting electrode. Thus, the chip will be
mainly heated and achieve an effective temperature greater that the
bath temperature, since we are continuously pumping energy into the
system. To verify that the thermal effects dominate the radiation
power dependence of switching current, we moved the fiber to radiate
directly on a small area of the bare substrate, which is about $0.7$
mm away from the junction area. We then performed the same switching
current measurements at $16$ mK and obtained the radiation power
dependence of $R_{b}$, shown in Fig.~3 (black circles). It is shown
that the photon responsivity is apparently weaker, when radiating on
the bare substrate of a certain distance away from the junction than
that when focusing on the junction area. This is a reasonable
result, which can be attributed to a nonuniform temperature
distribution around the irradiated area. The effective temperature
at the junction area is lower when the light spot is moved $0.7$ mm
away. The $R_{b}$ exhibits the same radiation power law dependence
(the same slope) for both cases of radiation on the junction and the
substrate, indicating that the thermal effect is the main factor in
shifting the switching current.
\begin{figure}[tbp]
\begin{center}
\begin{overpic}[width=7.5cm]{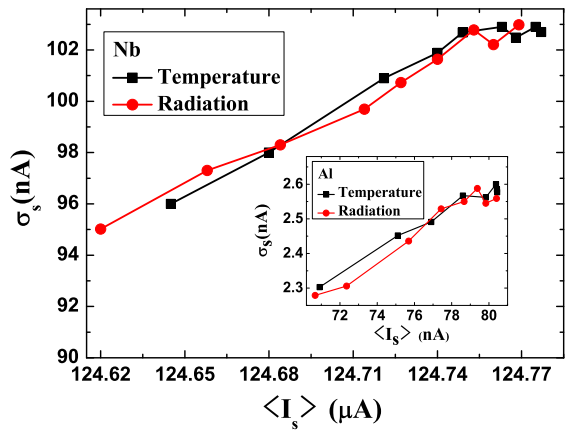}
\end{overpic}
\end{center}
\small {\caption{(Color online) The list plots of the measured
distribution width $\sigma_{s}$ versus the average switching current
$\langle I_{s}\rangle $, for the Nb-junction, due to independent
changes in bath temperatures (black squares) and radiation powers
(red circles) respectively. The inset shows the same plots for the
Al-junction.}}
\end{figure}

Furthermore, we measured the variations of the switching current
distribution due to the changes in bath temperature and radiation
power independently. Fig.~4 plots the distribution width
$\sigma_{s}$ as a function of the average switching current $\langle
I_{s}\rangle $ for the Nb device. Here, the black squares correspond
to the data at different bath temperatures and without radiation,
while the red circles correspond to different radiation powers and
at the lowest bath temperature. The inset shows the same plots for
the Al device. It is seen that for both Nb and Al devices,
$\sigma_{s}$ approximately follows the same function of $\langle
I_{s}\rangle$ by varying the bath temperatures or radiation powers:
the distribution width has a plateau in higher switching current
regime and then decreases with decreasing switching current
monotonically~\cite{yuhaifeng}. In another word, one can obtain a
certain switching current distribution $P(I_{s})$ by radiating the
junction with a certain power at a fixed bath temperature, and the
same $P(I_{s})$ (i.e., the same $\sigma_{s}$ and $\langle
I_{s}\rangle$) can also be obtained with an un-irradiated junction
by setting the bath temperature at a certain value. This suggests
again that for the present devices, the optical responses of the
switching current are mainly due to thermal effects and therefore
the present junctions can be used as a desirable bolometer.

Principally, the weak light detection scheme in time domain is
straightforward. One can bias the junction at a current slightly
smaller than its switching current in the absence of radiation. If a
light pulse with sufficient energy is applied, the switching current
of the junction is reduced to below the bias current and then the
junction will switch to a finite voltage state. Otherwise the
junction will stay in zero voltage state. In this way one can judge
if there are incoming photons or not.

Although the photon responsivity of the junction device demonstrated
here is obviously lower than that of other superconducting detectors
(such as the TESs and MKIDs), its performance in the weak light
detection could be further improved by several methods. The first
one is to enhance the coupling between the superconducting electrode
of the junction and the incident photons. To this aim, one can use
the lensed fibers to focus the light on the top electrode so that a
maximum energy from the incident photons could be absorbed directly
by the Cooper pairs on the electrode. Besides, the top metal
electrode can be fabricated as thin as possible so that a certain
radiation power can lead to a more reduction in the Cooper pair
density. The second method to raise reponsivity is to reduce the
thermal conductance between the chip and the sample holder to
maximize the energy absorption by the whole chip. For instance, one
can fabricate the junction on the substrate with low thermal
conductivity (e.g., amorphous glass) or etch the back of the
substrate wafer, to reduce the path of heat conduction from the chip
to the sample block. By this way unnecessary radiation energy loss
can be avoided effectively. The third method is to select materials
with lower gap energies as the superconducting electrodes. Finally,
for our experiments the fluctuations in the mean switching currents
are limited by the low-frequency noises in the circuits, not by
intrinsic noises of the junctions. Therefore, the device performance
can be potentially improved by further reducing the noise in the
measurement electronics.

Besides the optical responsivity, there are two more challenges in
our studied detection system. Firstly, the detection mechanism
(i.e., by measuring the switching currents of Josephson junctions)
can not be made very fast, since it takes time to sweep up the
current and then sweep down to reset the junction to zero voltage
state. The present measurement can be done at the rate up to several
KHz, which is still a lower rate. By improving the bandwidth of the
measurement electronics, the measurement rate can be faster but may
not easy to get up to MHz. In addition, a faster measurement (i.e.,
a faster current ramping rate) will broaden the distribution width
of the switching currents, which can decrease the detection
sensitivity. Secondly, thermal activation is greatly suppressed at
ultra-low bath temperatures, but the distribution of the switching
currents still has a finite width $\sigma_{q}$ due to quantum
tunneling. For our tested Nb-junction sample and experimental
parameters, one can calculate~\cite{cal} $\sigma_{q}$ = $101.87$ nA,
which is very close to the observed distribution width
$\sigma_{s0}(Nb)$ = $102.73$ nA at $T =$ $16$ mK, indicating the
quantum tunneling is dominant at ultra-low bath temperatures. This
finite distribution width may be translated to high dark counts for
photon counting (as shown in Fig.~2, the blue and red curves have
overlaps). However, this problem disappears when the junction device
is only used as an optical power meter. One can statistically
average the switching currents to distinguish the incident light
powers. In the case of weak low-frequency circuit noise, the
fluctuations in the average switching currents could be very small
and thus the junction device can be utilized as a very sensitive
radiation power meter.

In summary, we experimentally investigated the optical responses of
the switching currents for the Al/AlOx/Al and the Nb/AlOx/Nb
Josephson junctions at ultra-low temperatures ($T$ $\approx$ $16$
mK). The radiation power dependence of the relative switching
current shifts were measured for both junctions. It was found that
the Al-junction has a more sensitive optical response than the
Nb-junction. The minimum radiation powers that the Al and Nb devices
can respond to are about $8$ pW and $2$ nW respectively. Moreover,
the Al-junction has been irradiated directly and indirectly through
the substrate. It was observed that, the relative switching current
shifts for both cases follows the same radiation power law
dependence, indicating that the thermal effects are dominant in the
optical responses. Hopefully, the junction devices demonstrated here
can be applied to implement photon detections in the future, once
the photon responsivity can be further improved.

\vspace{0.4cm}

This work was supported in part by the National Natural Science
Foundation (Grant Nos. 61301031, 61371036, 11174373, 11204249), the
Fundamental Research Funds for the Central Universities (Grant No.
2682014CX087) and the National Fundamental Research Program of China
(Grant No. 2010CB923104). We thank Profs. Peiheng Wu, Xuedong Hu and
Yang Yu for kind supports and valuable discussions.

\nocite{*}
\bibliography{aipsamp}

\begin{thebibliography}{1}
\bibitem{G} G. N. Gol'tsman, O. Okunev, G. Chulkova, A. Lipatov, A. Semenov, K. Smirnov, B. Voronov, A. Dzardanov, C. Williams and R. Sobolewski, App. Phys. Lett. \textbf{79}, 705 (2001).
\bibitem{H} H. Kesue, S. W. Nm, Q. Zhng, R. H. Hdfield, T. Honjo, K. Tmki and Y. Ymmoto, Nature Photonics \textbf{1}, (343) 2007.
\bibitem{A} Aleksander D., Francesco M., David B., et al., Nature Photonics \textbf{2}, (302) 2008.
\bibitem{AA} F. Marsili, V. B. Verma, J. A. Stern, S. Harrington, A. E. Lita, T. Gerrits, I. Vayshenker, B. Baek, M. D. Shaw, R. P. Mirin and S. W. Nam, Nature Photonics \textbf{7}, (210) 2013.
\bibitem{Miller} A. J. Miller, S. W. Nam, J. M. Martinis and A. V. Sergienko, App. Phys. Lett. \textbf{83}, 791 (2003).
\bibitem{Miller2} L. Adriana E., A. J. Miller, and S. W. Nam, Optics Express \textbf{16(5)}, 3032 (2008).
\bibitem{SF} S. Friedrich, M. H. Carpenter, O. B. Drury, W. K. Warburton, J.Harris, J. Hall and R. Cantor, J. Low Temp. Phys. \textbf{167}, (741) 2012.
\bibitem{JD} J. D. Teufel, Ph.D. Thesis, Yale University, 2008.
\bibitem{pkday1} P. K. Day, H. G. LeDuc, B. A. Mazin, A. Vayonakis, and J. Zmuidzinas, Nature(London) \textbf{425}, 817 (2003).
\bibitem{gao1} J. Gao, M. R. Visser, M. O. Sandberg, F. C. S. da Silva, S. W. Nam, D. P. Pappas, D. S. Wisbey, E. C. Langman, S. R. Meeker, B. A. Mazin, H. G. Leduc, J. Zmuidzinas, and K. D. Irwin, App. Phys. Lett. \textbf{101}, 142602 (2012).
\bibitem{gao2} J. Gao, Ph.D. thesis, Caltech, 2008.
\bibitem{feynman} R. P. Feynman, B. L. Robert, and L. S. Matthew, The Feynman Lectures on Physics, Vol. \textbf{3}, Basic Books, 2011.
\bibitem{AB} V. Ambegaokar and A. Baratoff, Phys. Rev. Lett. \textbf{10}, 486 (1963).
\bibitem{shapiro} C. C. Grimes, P. L. Richards, and Sidney Shapiro, J. App. Phys. \textbf{39(8)}, 3905 (1968).
\bibitem{Osterman1} D. P. Osterman, M. Radparvar, and S. M. Faris, IEEE Trans. Magn. \textbf{25(2)}, 1319 (1989).
\bibitem{Osterman2} D. P. Osterman, P. Marr, H. Dang, C-T. Yao, and M. Radparvar, IEEE Trans. Magn. \textbf{27(2)}, 2681 (1991).
\bibitem{wire} M. S. Wire, L. O. Heflinger, B. J. Dalrymple, M. Leung, T. Pham, L. R. Eaton, and A. H. Silver, IEEE Trans. App. Superconduct. \textbf{3(1)}, 2107 (1993).
\bibitem{Mont1} E. Monticone, V. Lacquaniti, R. Steni, M. Rajteri, M. L. Rastello, L. Parlato, and G. Ammendola, IEEE Trans. App. Superconduct. \textbf{9(2)}, 3866 (1999).
\bibitem{Mont2} E. Monticone, M. Rajteri, R. Steni, M. L. Rastello, V. Lacquaniti, G. P. Pepe, L. Parlato, and G. Ammendola, Int. J. Mod. Phys. B. \textbf{13(09n10)}, 1283 (1998).
\bibitem{Mont3} M. Rajteri, E. Monticone, G. B. Picotto, R. Steni, M. Rastello, and V. Lacquaniti, in Proceedings ICEC 17, D. Dew-Hughes, R. G. Scurlock and J. H. P. Watson, Eds. Bristol and Phyladelphia: IOP Publishing, 711 (1998).
\bibitem{Stewart} W. C. Stewart, App. Phys. Lett. \textbf{12}, 277 (1968).
\bibitem{Fulton} T. A. Fulton and L. N. Dunkleberger, Phys. Rev. B \textbf{9}, 4760 (1974).
\bibitem{Barone} A. Barone, R. Cristiano, and P. Silvestrini, J. App. Phys. \textbf{58}, 3822 (1985).
\bibitem{yiwen} G. Z. Sun, Y. W. Wang, J. Y. Cao, J. Chen, Z. M. Ji, L. Kang, W. W. Xu, Y. Yang, S. Y. Han, and P. H. Wu, Phys. Rev. B 77, 104531 (2008)
\bibitem{yuhaifeng} H. F. Yu, X. B. Zhu, Z. H. Peng, Y. Tian, D. J. Cui, G. H. Chen, D. N. Zheng, X. N. Jing, L. Lu, and S. P. Zhao, Phys. Rev. Lett. \textbf{107}, 067004 (2011).
\bibitem{cal} The theoretical distribution of switching currents due to quantum tunnelling is given
by the equation:
$P(I_s)=\frac{\Gamma_{q}(I_s)}{dI/dt}\exp\left[-\frac{1}{dI/dt}\int_0^{I_s}\Gamma_{q}(I)dI\right]$,
where $dI/dt$ $=$ $1.63\times 10^{-2}A/s$ is the bias current ramp
rate and $\Gamma_{q}(I)$ is the well-known macroscopic quantum
tunnelling (MQT) rate at bias current $I$. To calculate
$\Gamma_{q}$, one needs to know the critical current $I_{c}$ and
shunt capacitance $C$ of the Nb-junction. In our experiment, we
obtain $I_{c}$ $=$ $126.66$ $\mu$A by fitting to the average
switching current and $C$ $=$ $0.26$ pF from independent microwave
resonance experiment. The distribution width $\sigma_{q}$ = $101.87$
nA can then be extracted once the distribution $P(I_s)$ is obtained.


\end{thebibliography}

\end{document}